# Comparative molecular dynamics simulations of charged solid-liquid interfaces with different water models

Mahdi Tavakol[1,*] and Kislon Voïtchovsky[1,*]

1. Physics Department, Durham University, Durham DH1 3LE, UK
*corresponding authors: mahditavakol90@gmail.com, kislon.voitchovsky@durham.ac.uk

## Abstract

Aqueous solid-liquid interfaces (SLI) are ubiquitous in nature and technology, often hosting molecular-level processes with macroscopic consequences. Molecular dynamics (MD) simulations offer a tool of choice to investigate interfacial phenomena with atomistic precision, but there exists a large number of water models, each optimised for a different purpose. Here we compare the ability of common water models to accurately simulate the interface between a charged silica surface and an aqueous solution containing NaCl. We first compare the bulk dielectric constant of water and its dependence on salt concentration for SPC/Fw, SPC/e, TIPS3p, $H_2O$/DC, TIP3P-Fw, OPC3, TIP3P, TIP3P-FB, TIP3P-ST, FBA/e, and TIPS3p-PPPM, revealing large variations between models. Simulating the interface with silica for the most suitable water models (SPC/Fw, $H_2O$/DC, TIP3-ST and TIPS3p-PPPM) show some intrinsic consistency with continuum predictions (Poisson-Boltzmann) whereby the free energy minima obtained from MD and the analytical model are in agreement, provided the latter includes the MD-determined total charge of ions in the Stern layer and dielectric constant. This consistency stands even for water models with a dielectric constant off by 100%. For salt concentrations higher than 0.21 M NaCl, the formation of random ion-ion pairs limits the reproducibility of the MD results and the applicability of the analytical method. The results highlight the applicability of the analytical model down to the nanoscale, provided *a priory* knowledge of the Stern layer charge is available. The findings could have significant implications for MD simulations of SLIs, especially at charged or electrified interfaces.

# Introduction

Aqueous solid-liquid interfaces (SLI) are ubiquitous in nature, industry and technology, for example in minerals growth[1,2], the survival and bioenergetics of living organisms[3,4], electrochemical processes and energy storage[5], sensing[6], water purification[7] and in colloidal stability[8].

In all cases, ions are present in the water, even if only from the natural dissociation of water molecules. Most systems contain additional metal ions which play a significant role, especially for processes taking at the interface with contacting solids.

Several continuum analytical models describing the behaviour of ions at SLIs have been developed and refined in the last century[9,10], and currently serve as a point of reference for most SLI-based systems. The solid usually carried an electrical potential, either naturally due to the presence of surface charges[11] or applied externally in technological applications[12]. This causes the ions in solution to accumulate near the SLI to ensure electroneutrality. Models often refer to the electrical double layer[9] (EDL) for the region near the solid where the concentration of counter-ions exceeds that of the bulk solution, with triple layer models[13] also taking into account some solvation aspect of the ions in contact with the solid. These models are remarkably successful for providing a quantitative description of ionic densities near SLIs[9,10,14], but usually cannot predict the lateral arrangement of ions along the interface since discrete molecular details are needed. Phenomena such as charge-charge correlations, charge inversion and overscreening are also difficult to capture under continuum assumptions[15,16].

If molecular details of the water and the ions near the SLI is necessary, studies often rely on advanced experimental techniques such as scattering techniques[17,18], sum frequency generation[19] and scanning probing methods[20–25], or computer simulations[26].

Computational methods are particularly well suited for investigating the short time- and length-scales of molecular processes at SLIs. Molecular dynamics (MD) simulations, in particular, are extensively used to study ions at SLIs[27–31], including under externally applied electrical potentials[32–34], and for comparison with experimental studies[22,24,35,36]. In MD simulations, the forces experienced by each atom is obtained from interatomic forcefields[37], with the trajectory of every atom tracked throughout the simulation. For simulations at SLIs, force fields are needed to describe both the solid surface and the liquid. Usually, the choice of forcefield for the surface is dictated by the known physical behaviour of the solid with limited options. However, things are more complicated for the liquid with a vast variety of water models and forcefields.

The available water models can be grouped into different categories based on their resolution or their flexibility. The common water models, based on their resolution, can be categorized into 3-site[38–41], 4-site[40,42–48], 5-site[49–51] and 6-site[52,53] water models. In 3-site water models only the water atoms are considered[38–41], while adding a dummy atom with a negative charge to the bisector of the HOH angle brings about the 4-site model[40,42–48]. The 5-site water model has two dummy atoms alongside with the three atoms of the water molecules[49–51] and finally all the dummy atoms of the lower resolution water models are present in the 6-site model[52,53]. Also, depending on how the motion of water atoms are integrated in MD simulation, the water models can be either rigid or flexible[40,54–56]. The models are usually calibrated to reproduce selected known water properties[40,41,54–59], with its bulk density and dielectric constant as the most common points of reference.

Correctly capturing the dielectric constant of water is particularly important for MD simulations of SLIs since the constant modulates ionic interactions in an aqueous media. Ongoing optimisation efforts[54,55,57,60,61] have improved the models, but there remain important challenges. For example, the salt-induced reduction of the dielectric constant is well established[62–65], but the impact on the strength of ion-ion correlations at SLIs[22,34] is not clearly understood, nor is the salt concentration limit for the applicability of each water model. While various water models are routinely applied to describe SLIs[22,24,35,66,67], there does not seem to be a general consensus or set of guidelines to select the most suitable model for a given SLI.

In this study we systematically compare MD simulations obtained with common water models of the SLI between saline aqueous solutions and silica. We selected silica as a solid for its ubiquity in numerous natural[68] and technological systems[69,70], including for SLIs investigation[34,71]. We expect analytical predictions[9,10] to correctly describe the ionic density near the SLI for this system, hence providing a point of reference for comparison with MD simulations. The goal is to quantify similarities and differences between the predictions obtained from the different water models and establish the range of applicability in each case. We start by exploring the dielectric constant of pure and saline bulk water derived from each model.

## Methods

### System setup

Two different systems were utilized to compare water models in the presence of ions and at SLIs: (i) bulk water with various salt concentrations to obtain the variation in dielectric constant (Fig. 1a) and (ii) a saline aqueous solution near a charged silica surface to investigate the ionic distribution at the SLI (Fig. 1b). For bulk water simulations with each of the water models of SPC/Fw[54], SPC/e[41], TIPS3p[72], H$_2$O/DC[57], TIP3P-Fw, OPC3[58], TIP3P[40], TIP3P-FB[59], TIP3P-ST[55], FBA/e[56], TIPS3p-PPPM[72], 3000 molecules were placed inside a box with dimensions of $5 \times 5 \times 5$ nm$^3$ (see Table 1 for the associated model parameters). Subsequently, the effect of added salt on the water dielectric constant was investigated for four of the models, namely SPC/Fw, H$_2$O/DC, TIP3-ST and TIPS3p-PPPM. In each case, the same simulation box was used with six different NaCl concentrations: 0 M, 0.5 M, 1 M, 1.5 M, 2 M and 2.5 M.

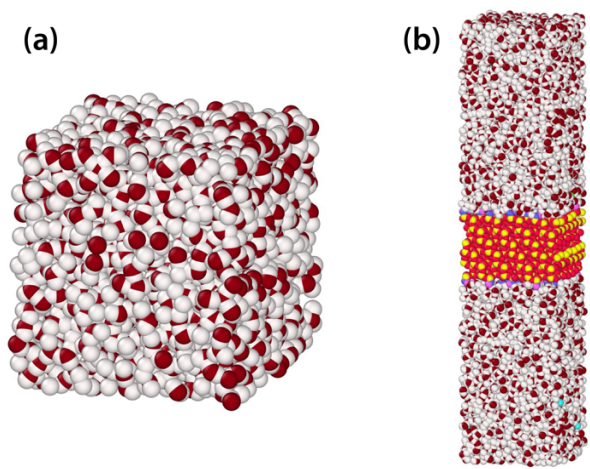

*Figure 1: System setup for the current study. The simulation box is shown for the bulk water simulations (a) and the simulation of the SLI between water and silica (b). Added ions are not shown for clarity.*

The same four water models were utilized to study the ionic distribution near a charged silica surface and compare the derived ionic distribution with predictions derived from the Poisson-Boltzmann theory, specifically the Gouy-Chapman[9,73,74] analytical SLI model. The silica Q3 surface was taken from the interface forcefield database[75], with the assumption of a pH of 7 and 0.67 SiO$^-$Na$^+$ groups per nm$^2$. The cross-sectional dimension of the simulation box was chosen equal to the cross section of the silica (3.7 $\times$ 3.7 nm$^2$) to assure an infinite silica with periodic boundary conditions while the lateral size of 16.5 nm for the box was selected to avoid the interaction of the silica slab of 2.3 nm thickness with its periodic image.

All the simulations were conducted in several steps: (i) water and ions were randomly placed inside the simulation box, followed by (ii) an energy minimization and subsequently (iii) several short equilibration stages before the (iv) main simulation which ran for 50 ns for bulk water, and 30 ns for dielectric calculations and SLI simulations. The temperature and pressure were kept constant at 300 K and 1 atm using a Nose-Hoover thermostat and barostat. The electrostatic interactions were separated to short-range and long-range interactions with the latter been calculated through Particle-Particle Particle-Mesh (PPPM) algorithms[76] (see Table 1). The non-bonded and short-range electrostatic interactions were cutoff with the CHARMM potential cutoff function in which the force and energy smoothly reaches zero in the region between inner and outer radius values[77]. All the simulations were done with the 23 Jun 2022 version of LAMMPS[78]. OVITO[79] and Matplotlib package of the python[80] were used for the visualisation and several C++ and python codes were developed for analysis of the results.

*Table 1: List of the different water models used in the study, together with the relevant simulation parameters and literature reference.*

| Model | $\sigma_O$ | $\varepsilon_O$ | $\sigma_H$ | $\varepsilon_H$ | $q_H$ | $k_{OH}/2$ | $r_{OH}$ | $K_\theta$ | $\theta$ | Flexible? | LJ/Coul | LJ/Coul | PPPM | Ref |
|---|---|---|---|---|---|---|---|---|---|---|---|---|---|---|
| SPC/Fw | 3.16549 | 0.155425 | 0 | 0 | 0.41 | 529.581 | 1.012 | 37.95 | 113.24 | Yes | 9 | 9 | Yes | [43] |
| SPC/e | 3.166 | 0.1553 | 0 | 0 | 0.4238 | --- | 1 | --- | 109.47 | No | 9 | 9 | Yes | [46] |
| TIPS3P | 3.1507 | 0.1521 | 1.7753 (0.4) | 0.0836 (0.046) | 0.4170 | --- | 0.9584 | --- | 104.5 | No | 10 | 12 | No | [47] |
| $H_2O$/DC | 3.1840 | 0.14173 | 0 | 0 | 0.45495 | --- | 0.958 | --- | 109.471 | No | 8 | 10 (12) | Yes | [48] |
| TIP3P-Fw | 3.188 | 0.102 | 0 | 0 | 0.415 | 529.581 | 0.9572 | 34.04355 | 104.52 | Yes | 10 | 12 | Yes | [29] |
| OPC3 | 3.17427 | 0.163406 | 0 | 0 | 0.447585 | --- | 0.97888 | --- | 109.47 | No | 8 | 8 | Yes | [49] |
| TIP3P | 3.188 | 0.102 | 0 | 0 | 0.415 | --- | 0.9572 | --- | 104.52 | No | 13 | 13 | Yes | [29] |
| TIP3P- | 3.178 | 0.155 | 0 | 0 | 0.424 | --- | 1.011 | --- | 108. | No | 9 | 7 | Yes | [50] |

| | | | | | | | | | | | | | |
|---|---|---|---|---|---|---|---|---|---|---|---|---|---|
| FB | | 86 | | | 22 | | 8 | | 15 | | | | |
| TIP3P-ST | 3.19257 | 0.143858 | 0 | 0 | 0.42556 | --- | 1.02030 | ---- | 108.11 | No | 9 | 9 | Yes | [44] |
| FBA/e | 3.1776 | 0.18937 | 0 | 0 | 0.4225 | 358.509 | 1.024 | 45.7696 | 114.7 | Yes | 9 | 9 | Yes | [45] |
| TIPS3p-PPPM | 3.1507 | 0.1521 | 1.7753 (0.4) | 0.0836 (0.046) | 0.4170 | --- | 0.9584 | --- | 104.5 | No | 10 | 12 | Yes | [47] |

**Dielectric constant calculations**

The water dielectric constant $\varepsilon$ for each water models was calculated from a simulated water box with the desired salt concentration (Fig. 1a) and using equation 1, where $M$, $\varepsilon_0$, $V$, $k_B$, $T$ and $\langle \ \rangle$ represent the total electric dipole, vacuum permittivity, volume, Boltzmann constant, temperature and ensemble average. A dedicated C++ code was written to calculate the electric dipole (eq. 1). The ensemble averages for the $M$ and $M^2$ parameters were calculated from their variation over the course of the simulation (equations 2-3).

$$\varepsilon = 1 + \frac{\langle M^2 \rangle - \langle M \rangle^2}{3\varepsilon_0 V k_b T} \tag{1}$$

$$\langle M \rangle = \frac{\int_0^t M \, dt}{t} \tag{2}$$

$$\langle M^2 \rangle = \frac{\int_0^t M^2 \, dt}{t} \tag{3}$$

**The Gouy-Chapman model**

The MD simulation results for the silica slab in contact with four different water models (Fig Fig. 1b) were compared with Poisson-Boltzmann predictions through the Gouy-Chapman model[9,73,74] as described by equations 4-6. In this model, the ions balance electrostatic interaction with the silica surface (Poisson equation[9]) with entropic considerations through the Boltzmann distribution. For a planar interface[9], this leads to an exponential decrease (respectively increase) in the concentration of counter-ions (co-ions) from the surface of the solid to the bulk ionic concentration. The characteristic length scale of the exponential evolution is characterized by the so-called Debye length $\lambda$ (equ. (4)) which depends on $\varepsilon$, $\varepsilon_0$, $k_B$, $T$, the electron charge $e$, the Avogadro number $N_A$ and the bulk concentration $c_0$ of salt in the system. Here, since a monovalent salt was used, $c_0$ represents the concentration for both anion and cations. The value of $\varepsilon$ was taken from the experimentally measured values by Buchner et al.[62] for different NaCl concentrations unless stated otherwise. The electrical potential $\psi_0^i$ at the surface of silica and the potential distribution $\psi^i(z)$ along the direction normal to the surface are obtained through equations 5 and 6 where $\sigma^i$ is the surface charge density, $z - z_0^i$ the distance from the surface and the index $i$ refers to either of the two surfaces of the silica slab. Finally, the density $\rho^i(z)$ of ions in the solution at any distance $z$ from the surface of the silica is obtained through equation 7 in which the positive and negative signs are used for cations and anions respectively.

$$\lambda = \sqrt{\frac{\varepsilon \varepsilon_0 kT}{2e^2 N_A * 1000 * c_0}} \tag{4}$$

$$\psi_0^i = -2\frac{kT}{e} \sinh^{-1}(\frac{\lambda e \sigma^i}{2\varepsilon \varepsilon_0 kT}) \tag{5}$$

$$\psi^i(z) = 4\frac{kT}{e} \tanh^{-1}(\tanh\left(\frac{e\psi_0^i}{kT}\right) * \exp(-\frac{z - z_0^i}{\lambda})) \tag{6}$$

$$\rho^i(z) = c_0 \exp(\frac{\pm e\psi_0^i}{kT}) \tag{7}$$

An ion condensation parameter $\varphi(z)$ is used to compare the MD results with the analytical model with $\varphi(z)$, defined through the equation (8) with $a$ is chosen as the distance where the counter-ion concentration reaches its maximum value near the surface in MD simulations.

$$\varphi(z) = \frac{-N_A e}{\sigma} \int_a^z (\rho(\acute{z}) - c_0) d\acute{z} \tag{8}$$

**Free energy calculations**

The Poisson-Boltzmann hypothesis underpinning the Gouy-Chapman model assumes that electrostatic interactions and entropy determine the free energy of the system (Equations 5 and 7). Here, the free energy was also calculated through MD simulations for comparison. The free energy of the process in which ions dissolved in the bulk solution (reactant) are adsorbed at the surface (products) is calculated from equation (9) where $\Delta G$ represents the change in free energy associated with the process. $\Delta G^0(z)$ is the standard free energy, the amount of free energy required to convert (move) a mole of the reactant (bulk ions) to the product (ions at distance z from the surface). Since the ions adsorbed on the surface are at equilibrium with those in the bulk, the free energy of the reaction is zero yielding equations 10 and 11. Equations 11 and 7 rest on the Poisson-Boltzmann assumption and are hence similar. Here, comparing the free energy $G(z)$ obtained from MD simulation with the $e\psi_0^i$ calculated from the analytical solution is used to test the Poisson-Boltzmann hypothesis by ignoring potential energies other than the electrostatic interaction in the Boltzmann distribution.

$$\Delta G(z) = \Delta G^0(z) + kT \ln\left(\frac{\rho^i(z)}{c_0}\right) \tag{9}$$

$$\Delta G^0(z) = G(z) = kT \ln\left(\frac{\rho^i(z)}{c_0}\right) \tag{10}$$

$$\rho^i(z) = c_0 \exp\left(\frac{G(z)}{kT}\right) \tag{11}$$

## Results

**Salt dependence of the dielectric constant for different water models**

Results from the simulations show that the derived dielectric constant of bulk water strongly depends on the water model considered (Fig. 2a), with the flexible TIP3p model yielding the highest value ($\varepsilon = 183.05 \pm 27.66$), more than twice the experimental value of 78 for pure water at the same temperature[81]. The TIPS3p water model which is mainly used with the CHARMM forcefield for biological systems [69] also overestimates the dielectric constant ($\varepsilon = 104.68 \pm 3.98$) but the application of the PPPM algorithm for the calculation of long-range electrostatic interaction slightly reduces the value to $\varepsilon = 97.66 \pm 4.80$.

Among the different water models considered here, SPC/Fw, $H_2O$/DC, TIP3P-ST, OPC3 and FBA/e yield values closet to the known water dielectric constant. We therefore selected the SPC/Fw, H2O/DC and TIP3P-ST water models to investigate the effect of salt concentrations on their dielectric constant (Fig. 2b). We also investigated the TIPS3p-PPPM water model due to its widespread in the field. As expected, the dielectric constant reduces with increasing the salt concentration of the solution for all the models investigated here. The SPC/Fw model correctly replicates the experimental values up to a concentration of 0.5 M but deviates for larger concentrations. The TIP3P-ST and H2O/DC models overestimate $\varepsilon$ for concentrations below 0.5 M and predict similar values as the SPC/Fw water model at higher salt concentrations. The TIPS3p-PPPM water model consistently overestimates $\varepsilon$ by the largest amount at all concentrations, with its prediction at 2.5 M NaCl almost twice the experimental value of $\varepsilon = 44$.

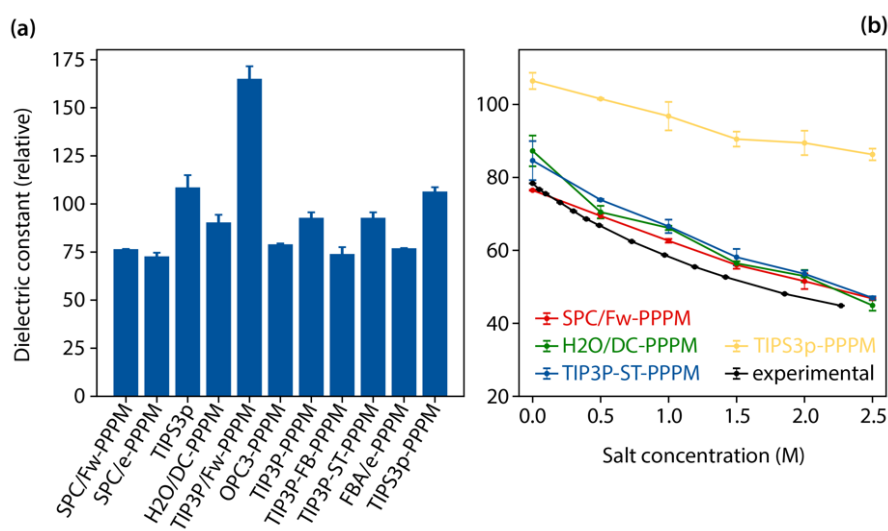

*Figure 2*: *Dielectric constants of bulk aqueous solutions calculated from MD simulations with different water models. In pure water (a), the 11 different water models exhibit significant differences. The three models best reproducing the experimental dielectric constant (SPC/Fw, H2O/DC and TIP3P-ST), together with the widely used TIPS3p-PPPM are selected to further calculate the effect of added NaCl (b). The models correctly capture the decrease in dielectric constant with increasing salt concentration, but they all tend to overestimate the value, with SPC/Fw-PPPM performing best at lower salt concentration (<0.5 M). For comparison, the experimental values[62] are also shown.*

The significant deviations between simulated and experimental dielectric constant raises obvious questions regarding their impact on the simulated ionic distribution at interfaces, here with the silica. These questions are addressed in the subsequent section.

**The solid-liquid interface for different water models**

The distribution of cations and anions near a negatively charged silica surface confirm that all the water models qualitatively capture the expected behaviour of the EDL near the surface (supplementary Fig. S1). All show a gradual decrease in the concentration of co-ions concomitant with an increase in the concentration of counter-ions near the charged silica. Decreasing the bulk salt concentration increases the distance needed for the ions' concentration to reach the bulk concentration (Debye length), also in line with analytical predictions.

Using the counter-ion/co-ion density profiles in conjunction with equation 11, the free energy required to move a counter- or co-ion from bulk to the position (z) can be calculated (Fig. 3). The free energy profiles for the counter-ions exhibit a minimum whose value decreases with increasing salt concentration for all water models (see supplementary Fig. S2 for the detailed profiles). However, the position of the minimum is consistently positioned 1.65-1.70 nm from the surface, regardless of the water model used or the salt concentration. This suggests the presence of dense counterions layer (Stern layer) which properties are dominated by the counterions rather than the water molecules. Here, the Stern layer is defined as the region located between the free energy minima and the silica surface. The total charge of counter-ions and co-ions in the Stern layer was calculated for comparison with silica's surface charge. The charge of the ions in each Stern layer increases with the salt concentration for all the water models. At the highest salt concentration (0.84 M), all the models converge to the same Stern layer total charge except for TIPS3P-PPPM (Fig. 3a). The charge, around +4.5e, is slightly higher than half the surface charge of silica (-8e for each side). This was taken into account for comparison with the Gouy-Chapman model with $\sigma^i$ taken as an effective total charge of the silica surface plus the Stern layer.

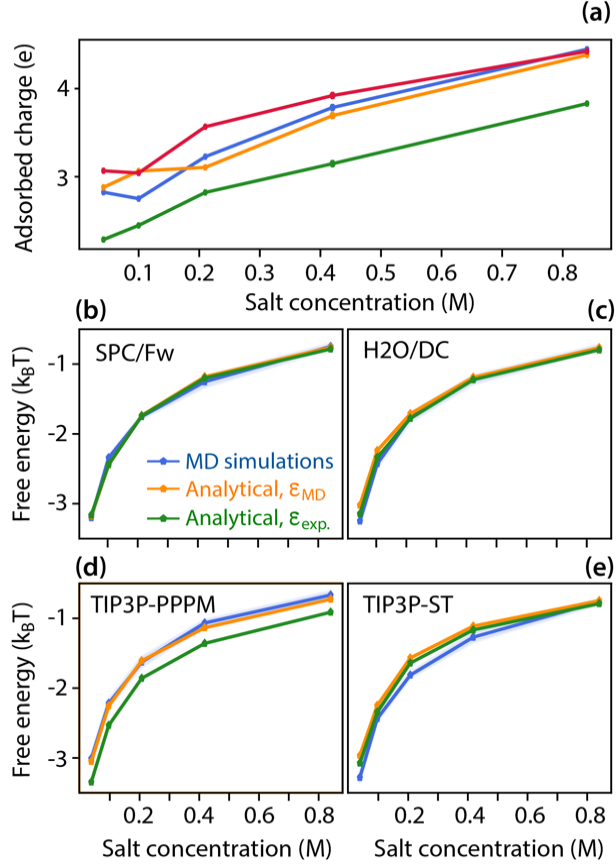

*Figure 3*: *Free energy of counter-ions near the silica-water interfaces obtained for different water models as a function of the solution's salt concentration. The total charge of counter- and co-ions obtained from MD simulations (a) show some disagreement between the models at lower salt concentration, with TIPS3P-PPPM consistently underestimating the value. The free energy of a counter-ion (cation) (b-e) is obtained through MD simulation using the electrostatic energy of the analytical model for SPC/Fw (b), H2O/DC (c), TIPS3P-PPPM (d), and TIP3P-ST (e). In each case, the analytical solution is given with using the dielectric constant derived from the MD ($\varepsilon_{MD}$) and experimentally measured ($\varepsilon_{exp}$.)[62] and the value simulation of the present study. The errors are shown as shaded around the reported values.*

To compare the MD simulation results with the analytical models, the free energy minimum obtained through the simulation (minimum of $G(z)$ from equation 10) was compared with minimum electrostatic potential energy ($e\psi_0^i$ from equation 5) of the analytical solution. Two sets of analytical equations were considered; the first with the reported water dielectric constant obtained from experimental measurements[62] and the second using the dielectric values obtained from each water model at the specific salt concentration considered (Fig 3b-e). Since two EDLs are present in each simulation, the value reported here is the average of the free energies from both the layers. Despite the variation in the dielectric constant (Fig. 2b) and charge of the Stern layer (Fig. 3a), the MD results exhibit a good agreement with the analytical predictions for all the water models when using the experimental dielectric constant (Fig. 3b-e) except for TIPS3P-PPPM. The agreement is better at higher salt concentrations. Predictions from the TIP3P-PPPM model (Fig. 3d) show a deviation in the range of 0.25 $k_BT$ - 0.33 $k_BT$ from the analytical value when using the experimental dielectric constant. Interestingly, if the dielectric constant obtained from simulations is used, no noticeable difference can be seen between the analytical solution and the MD predictions. This demonstrates some intrinsic consistency of the model, despite the large error on the dielectric constant obtained.

Perhaps more intuitive, the comparison between the density profiles of counter- and co-ion near the interface obtained through MD and through analytical predictions (Fig. 4) show a good agreement for all water models and salt concentrations, at least for larger distances from the interface (> ~3 Å) where no Stern layer is present. Closer to the surface, the analytical model anticipates a gradual decline in co-ions density, but MD results show a sudden and complete elimination of the co-ions, especially at higher concentration (see 0.84 M in Fig. 4a-d). This effect is less pronounced at lower salt concentrations. All the water models except for TIPS3P-PPPM overestimate the counter-ion density profiles near the surface at the lowest salt concentration of (0.042 M, Fig. 4). This tendency is progressively reversed with increasing the salt concentration (see e.g. Fig. 4a) with TIPS3P-PPPM consistently underestimating the counter-ion density close to the Stern layer (Fig. 4c, g, k, o, s).

Until here, the experimentally measured value of $\varepsilon$ at each salt concentration has been used in the analytical predictions. It is instructive to also consider analytical predictions obtained with the $\varepsilon$ value obtained from the MD simulations. Here just SPC/Fw and TIPS3p-PPPM water models are selected for the comparison; SPC/Fw is chosen for its more accurate presentation of the dielectric constant while the TIPS3p-PPPM is offers a good point of comparison for a dielectric constant significantly deviating from the experimentally measured value. The results, obtained for 0.042 M and 0.84 M NaCl (Supplementary Fig. S3), show that SPC/Fw slightly underestimates the counter-ion density near the silica surface at 0.84 M when using the MD calculated dielectric constant, but the TIPS3p-PPPM predictions are in a good agreement with the analytical solution despite its poor prediction of $\varepsilon$.

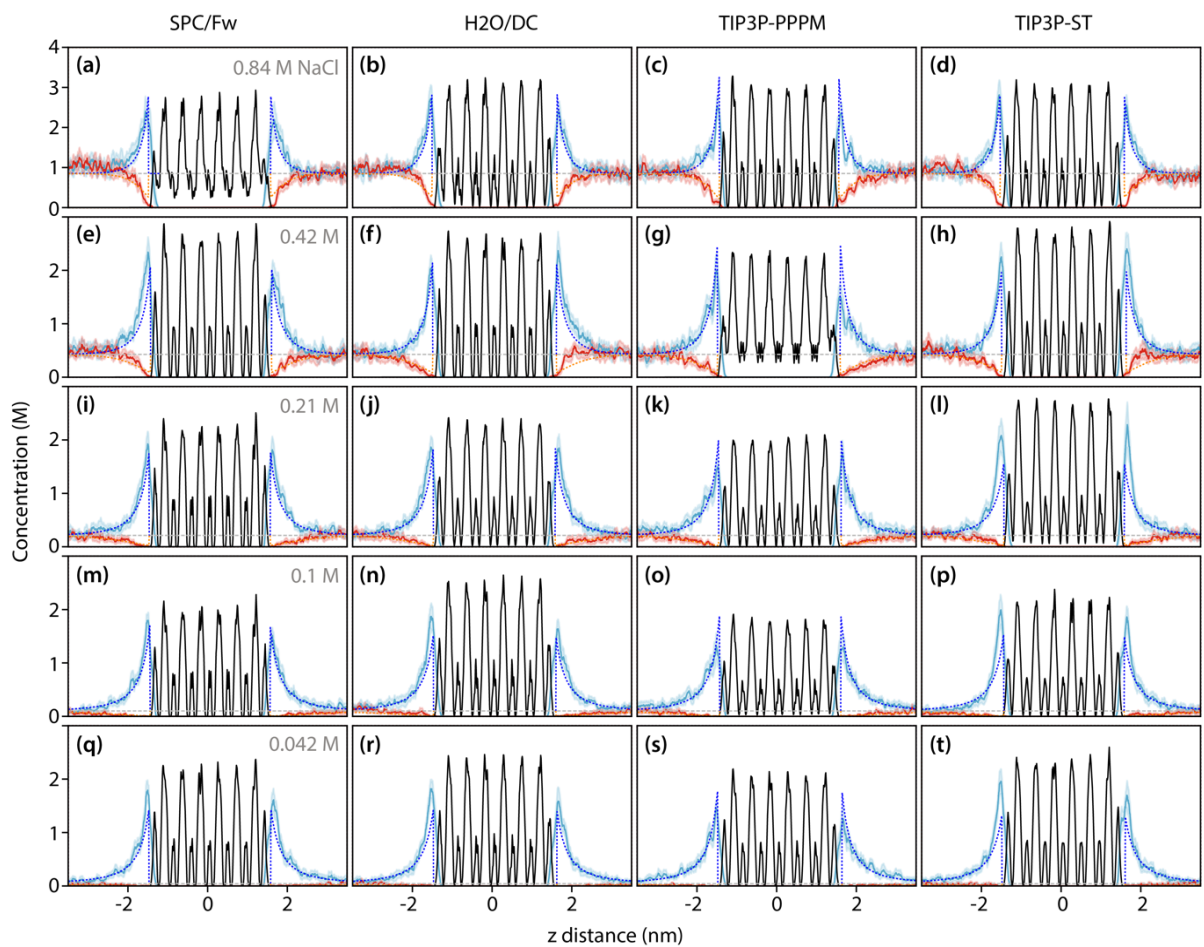

*Figure 4 - Comparison of the ion density profiles near the silica surface obtained with the four selected water models at various salt concentrations. In all cases, the simulated profiles are shown with solid lines and the equivalent analytical predictions with dashed lines. The profiles associated with the counter-ions (co-ions) appear blue (red). The SPC/Fw profiles at the bulk NaCl concentrations of 0.84 M, 0.42 M, 0.21 M, 0.1 M, and 0.042 are given in (a), (e), (i), (m), and (q), respectively. The $H_2O$/DC profiles at 0.84 M, 0.42 M, 0.21 M, 0.1 M, and 0.042 are given in (b), (f), (j), (n), and (r), respectively. The TIP3P-PPPM profiles at 0.84 M, 0.42 M, 0.21 M, 0.1 M, and 0.042 are given in (c), (g), (k), (o), and (s), respectively. TIP3P-ST profiles at 0.84 M, 0.42 M, 0.21 M, 0.1 M, and 0.042 are given in (d), (h), (l), (p), and (t), respectively.*

The condensation parameter values, as calculated through equation 8, are shown in Fig. 5 (see also Fig. S4 for more details). According to the Poisson-Boltzmann predictions, the condensation parameter should decrease (respectively increase) for co-ions (respectively counter-ions) when moving away from the surface of silica, with the condensation parameter converging to a fixed value beyond a specific distance, here between 2 and 3 Å from the surface (Fig. S4). For co-ions, this fixed value is systematically overestimated by MD, with a larger difference at higher salt concentration. The MD simulations tend to agree better with analytical predictions for the counter-ions up to concentrations 0.21 M (Fig. 4c-d, see also Fig. S3), except for the TIP3-ST model where the agreement stops beyond 0.1 M (Fig. S4p, t).

At higher salt concentrations the MD results overestimate the condensation parameter, as for co-ions. This similar behaviour for both counter and co-ions at distances larger than the Debye length suggests it is due to the formation of ion pairs in the simulations: the difference in the condensation profiles of counter- and co-ions remains fixed for both the SPC/Fw (Fig. 5a) and TIPS3p-PPPM (Fig. 5b). Counting the number of pairs in simulation snapshots further supports this interpretation with 37.25±0.23 pairs for for SPC/Fw and 34.49±0.98 for TIPS3p-PPPM at 0.84 M. In contrast, when the salt concentration is low enough to discourage the formation of ion pairs (e.g. 0.1 M yielding 0.26±0.04 and 0.53±0.02 pairs respectively), the two models tend to follow analytical predictions better (Fig. 4c-d).

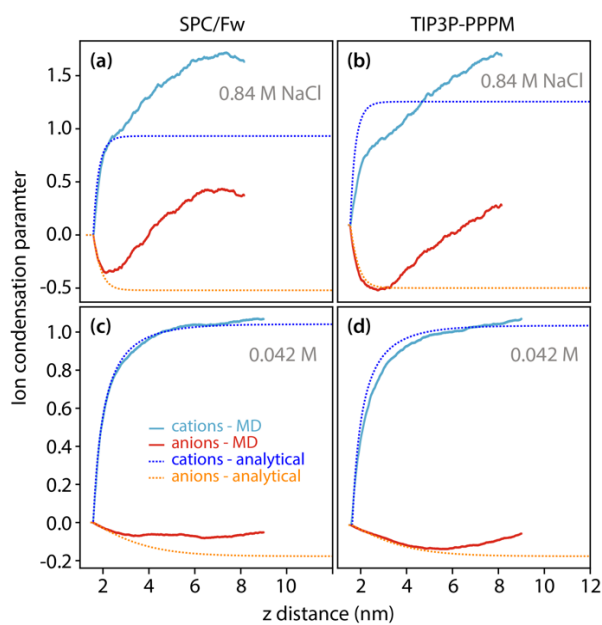

*Figure 5: Comparison of condensation parameter calculated analytically with the results from MD simulations with two water models: SPC/Fw and TIPS3P-PPPM. In each case, the results are given at 0.84 mM NaCl: SPC/Fw (a) and TIPS3P-PPPM (b), and at 0.042 M NacL: SPC/Fw (c) and TIPS3P-PPPM (d). The results are a subset of Fig. S3 and suggest the formation of ion pairs which are responsible for the differences between analytical and MD predictions.*

## Discussion

Solid-liquid interfaces are present in many natural and technological systems. The solid surface is usually either naturally charged or can carries an applied potential in technological applications, resulting in a reorganisation of the ions in the vicinity of the surface. MD simulations are routinely used to investigate such interfaces[33,66], often complementing and interpreting experiment results[22,24,34–36], but there is no clear consensus on the best choice of water model for a given application. Previous studies have reported efforts to fine-tune the dielectric constant of different water models[54,55,57–61], but little has been done to ensure a dielectric constant correctly reproducing the known evolution with salt concentration. Among the water models studied here, SPC/Fw[54], $H_2O$/DC[57], TIP3-ST[55] and TIPS3p-PPPM[72] can reproduce the experimentally measured dielectric constant of water up to a salt concentration of ~0.5 M, but they unilaterally overestimate the dielectric constant at higher salt concentrations. The higher accuracy observed for SPC/Fw at lower salt concentration might be due to the fact the associated $Na^+$ and $Cl^-$ parameters were taken from the values calculated in SPC/e water[82], a model very similar to SPC/Fw. Additionally, calibration of these parameters in SPC/e was conducted using the hydration free energy of ions[82], something particularly appropriate here. However, the TIPS3p-PPPM water model which, in combination with CHARMM forcefield family[83], is very common for the simulation of biological systems overestimated the dielectric constant over the whole salt concentration range probed despite the $Na^+$ and $Cl^-$ parameters taken from the CHARMM forcefield[84]. This is noteworthy considering the careful validation of the CHARMM family of forcefields and its widespread use in the literature[84–86], suggesting that some errors might cancel each other in the calculation of electrostatic interactions: even though the compatibility of ionic parameters with the water model is important, care should be taken to avoid situations in which errors cancel each other in the forcefield calibration.

When simulating the distribution of ions near the surface of charged silica, we show that the position of the minimum of free energy (Considered as the Stern layer) can be used to compared MD and analytical predictions. Using the nominal surface charge of the silica in the analytical model leads to an overestimate of the free energy with respect to the MD simulation for all the water models. However, including the ions of the Stern layer, defined as the region between the surface and the minimum of free energy, generally leads to a good agreement between the analytical and the MD results except for the TIP3P-PPPM water model.

Before discussing the TIP3P-PPPM simulations in more details, it is worth noting that the comparison between the other models and the analytical prediction is not straightforward. The analytical prediction uses experimentally measured dielectric constants and the charge of the Stern layer (for 0.042 M - 0.84 M salt concentrations). In contrast, the MD simulations rely on overestimated dielectric constants by up to 10, except for TIPS3P-PPPM model. The agreement between the analytical and MD free energy minimums leads to three observations. First, despite recent efforts to fine-tune the dielectric constant of some water models with tolerances of less than 1 [54,55,57–61], the exact value of the dielectric constant in the water model does not noticeably affect the free energy of ionic adsorption. This might be down to the weak dependence of the free energy on the dielectric constant ($\sqrt{\varepsilon}$ according to equations 4-5) or due to errors in MD simulations that are larger than the error introduced by the overestimation of the dielectric constant. So at least for the increase in the accuracy of the free energy values, which matters in interface driven phenomena, the fine-tuning of water models should direct towards other properties than the dielectric constant. Second, while the Gouy-Champan model can be applied here, there is a need for *a priori* knowledge of the Stern layer, at least for highly charged surfaces such as silica (0.08 C/m$^2$). Third, since the main contribution to the free energy is electrostatic interactions, ion-ion correlations do not appear significant for salt concentrations of up to 0.84M.

The predictions with the TIPS3-PPPM model tend to deviate from other models and analytical predictions. Replacing the experimentally measured dielectric constant with the TIPS3P-PPPM-deduced dielectric constant in the analytical predictions leads to a good agreement of the free energy between the MD and analytical results. This implies an opposite dependence of the Stern layer charge (and the total charge in equation 4 which is the sum of Stern charge and the surface charge) on the square root of the dielectric constant according to the equations 4 and 5. The higher dielectric constant of TIP3P-PPPM caused an over-screening of the electrostatic interactions, leading to lower counter-ion density near the surface. The observed agreement between MD simulations and PB theory with the water model dielectric constant even for TIP3P-PPPM demonstrates that MD reproduces the PB solution not because of a specific dielectric constant value, but because both enthalpic (electrostatic) and entropic (ionic distribution) contributions are intrinsically present in MD. This agreement holds particularly in the low-concentration regime, where ion–ion correlations and excluded volume effects are minimal, and validates the use of PB as a continuum approximation of the statistical behaviour captured atomistically in MD.

The simulated ionic distribution near the surface is generally in good agreement with the analytical prediction for salt concentrations of up to 0.21 M, except for the TIPS3P-PPPM. For higher salt concentrations care should be taken due to the formation of the ion-ion pairs not considered in the analytical model. The random nature of the ionic pair formation also increases errors in MD simulations. The fact that no clear difference could be observed between the ion distribution derived with the different water models (except for TIPS3P-PPPM) might seem surprising considering the different dielectric constants deduced here (79.83±2.16, 83.67±0.46 and 82.2±1.6 for SPC/Fw, H$_2$O/DC, TIP3-ST). Two possible reasons can be put forward: first, the uncertainty in the calculation of the ion distribution profile may dominate over a small (up to 10) change in the dielectric constant. Second, there might be a lack of convergence for the calculated dielectric constant, since it is derived through the equation 1 which converges slowly.

While insightful for SLIs simulations, the findings presented in this study are not devoid of limitations. First, we used $Na^+$ and $Cl^-$ calibrated with SPC/e water model for SPC/Fw, $H_2O$/DC and TIP3-ST water models due unavailability of the relevant parameter. This could be improved through a model-specific calibration of ion interaction parameters. The representation of the surface model is another limitation of the current study: the interface forcefield has been calibrated for the TIPS3p-PPPM forcefield and has cutoff values similar to the CHARMM forcefield. Switching the water model and the associated forcefield cutoff values changes the interface energy based on which the forcefield has been calibrated. However, since the focus of the current study is on the liquid part of the SLI, the silica surface can be taken as a generic charged surface, and its exact surface energy does not significantly affect the results. Finally, limited simulation time and length scale inevitably affects the accuracy of the results, something common to all MD studies. Despite these limitations, the comparative nature of the study offers novel insights into the use of different water models for MD simulations of aqueous SLIs, in particular their suitability and robustness for describing different aspects of the interface.

## Conclusion

In the present study, we compare the ability of different water models commonly used in MD simulations to correctly describe the interface between a charged solid surface and a saline aqueous solution. The results show that the choice of model can significantly affect the resulting dielectric constant of the saline solution, potentially also affecting the simulated ionic distribution at the SLI. Some water models correctly predict the variation in the dielectric constant with salt concentration in a dilute regime (<0.2 M), but all models tend to fail at higher salt concentrations likely due to ion-ion correlation effects. Comparison with the well-established Gouy-Chapman continuum formalism show a good general agreement for dilute salt concentration for several water models up to 0.42 M salt concentration but not for the widely adapted TIPS3p-PPPM. The current study sheds more light on the practical aspects of common water models in the context of MD investigations of aqueous SLIs. We anticipate the findings to be of interest for conducting simulations at charged or electrified interfaces.

## Associated content

Supporting Information

The Supporting Information is available free of charge at …

Simulated ionic profiles at the silica interface with different water models and salt concentrations (Fig. S1); associated free energy profiles (Fig. S2); comparison between MD and analytical predictions for ionic density with SPC/Fw and TIPS3p-PPPM (Fig. S3); ion condensation parameters at different salt concentrations (Fig S4).

**Our supporting research dataset is published in the Durham University research data repository. DOI:** (assigned upon manuscript acceptance)

## Acknowledgements

The authors wish to acknowledge the Hamilton supercomputer at Durham university for the provision of the computational facilities and support. This project has received funding from the UK Engineering and Physical Sciences Research Council (EPSRC grant EP/S028234/1).

## Authors contribution

MT and KV designed the problem. MT conducted the simulations and analysed the results with input from KV. MT and KV wrote the manuscript.

## Notes

The authors declare no competing financial interests.

## References

(1) Ricci, M.; Segura, J. J.; Erickson, B. W.; Fantner, G.; Stellacci, F.; Voitchovsky, K. Growth and Dissolution of Calcite in the Presence of Adsorbed Stearic Acid. *Langmuir* **2015**, *31* (27), 7563–7571.

(2) Weber, J.; Bracco, J. N.; Yuan, K.; Starchenko, V.; Stack, A. G. Studies of Mineral Nucleation and Growth Across Multiple Scales: Review of the Current State of Research Using the Example of Barite (BaSO4). *ACS Earth Space Chem.* **2021**, *5* (12), 3338–3361.

(3) Springer, A.; Hagen, V.; Cherepanov, D. A.; Antonenko, Y. N.; Pohl, P. Protons Migrate along Interfacial Water without Significant Contributions from Jumps between Ionizable Groups on the Membrane Surface. *Proc. Natl. Acad. Sci. U. S. A.* **2011**, *108* (35), 14461–14466.

(4) Coste, B.; Mathur, J.; Schmidt, M.; Earley, T. J.; Ranade, S.; Petrus, M. J.; Dubin, A. E.; Patapoutian, A. Piezo1 and Piezo2 Are Essential Components of Distinct Mechanically Activated Cation Channels. *Science* **2010**, *330* (6000), 55–60.

(5) Källquist, I.; Lindgren, F.; Lee, M.-T.; Shavorskiy, A.; Edström, K.; Rensmo, H.; Nyholm, L.; Maibach, J.; Hahlin, M. Probing Electrochemical Potential Differences over the Solid/Liquid Interface in Li-Ion Battery Model Systems. *ACS Appl. Mater. Interf.* **2021**, *13* (28), 32989–32996.

(6) Wachta, I.; Balasubramanian, K. Electroanalytical Strategies for Local pH Sensing at Solid–Liquid Interfaces and Biointerfaces. *ACS Sens.* **2024**, *9* (9), 4450–4468.

(7) Deng, D.; Aouad, W.; Braff, W. A.; Schlumpberger, S.; Suss, M. E.; Bazant, M. Z. Water Purification by Shock Electrodialysis: Deionization, Filtration, Separation, and Disinfection. *Desalination* **2015**, *357*, 77–83.

(8) von Smoluchowski, M. Zur Kinetischen Theorie Der Brownschen Molekularbewegung Und Der Suspensionen. *Annalen der Physik* **1906**, *326* (14), 756–780.

(9) Israelachvili, J. N. *Intermolecular and Surface Forces, 3rd ed.*; Academic press, 2011.

(10) Trewby, W.; Tavakol, M.; Jaques, Y. M.; Voïtchovsky, K. Towards Local Tracking of Solvated Metal Ions at Solid-Liquid Interfaces. *Mater. Today Phys.* **2024**, *44*, 101441.

(11) Sjöberg, S. Silica in Aqueous Environments. *J. Non-Cryst. Solids* **1996**, *196*, 51–57.

(12) Brett, C. M. A.; Brett, A. M. O.; Brett, C. M. A.; Brett, A. M. O. *Electrochemistry: Principles, Methods, and Applications*; Oxford University Press: Oxford, New York, 1993.

(13) Tournassat, C.; Chapron, Y.; Leroy, P.; Bizi, M.; Boulahya, F. Comparison of Molecular Dynamics Simulations with Triple Layer and Modified Gouy–Chapman Models in a 0.1 M NaCl–Montmorillonite System. *J. Coll. Interf. Sci.* **2009**, *339* (2), 533–541.

(14) Zhao, C.; Ebeling, D.; Siretanu, I.; van den Ende, D.; Mugele, F. Extracting Local Surface Charges and Charge Regulation Behavior from Atomic Force Microscopy Measurements at Heterogeneous Solid-Electrolyte Interfaces. *Nanoscale* **2015**, *7* (39), 16298–16311.

(15) Tan, Q.; Zhao, G.; Qiu, Y.; Kan, Y.; Ni, Z.; Chen, Y. Experimental Observation of the Ion–Ion Correlation Effects on Charge Inversion and Strong Adhesion between Mica Surfaces in Aqueous Electrolyte Solutions. *Langmuir* **2014**, *30* (36), 10845–10854.

(16) Calero, C.; Faraudo, J.; Bastos-González, D. Interaction of Monovalent Ions with Hydrophobic and Hydrophilic Colloids: Charge Inversion and Ionic Specificity. *J. Am. Chem. Soc.* **2011**, *133* (38), 15025–15035.

(17) Fenter, P.; Park, C.; Nagy, K.; Sturchio, N. Resonant Anomalous X-Ray Reflectivity as a Probe of Ion Adsorption at Solid-Liquid Interfaces. *Thin Sol. Films* **2007**, *515* (14), 5654–5659.

(18) Bourg, I. C.; Lee, S. S.; Fenter, P.; Tournassat, C. Stern Layer Structure and Energetics at Mica-Water Interfaces. *J. Phys. Chem. C* **2017**, *121* (17), 9402–9412.

(19) Pullanchery, S.; Kulik, S.; Rehl, B.; Hassanali, A.; Roke, S. Charge Transfer across C–H···O Hydrogen Bonds Stabilizes Oil Droplets in Water. *Science* **2021**, *374* (6573), 1366–1370.

(20) Fukuma, T.; Garcia, R. Atomic- and Molecular-Resolution Mapping of Solid-Liquid Interfaces by 3D Atomic Force Microscopy. *ACS Nano* **2018**, *12* (12), 11785–11797.

(21) Contera, S. A.; Voïtchovsky, K.; Ryan, J. F. Controlled Ionic Condensation at the Surface of a Native Extremophilemembrane. *Nanoscale* **2010**, *2* (2), 222–229.

(22) Ricci, M.; Spijker, P.; Voitchovsky, K. Water-Induced Correlation between Single Ions Imaged at the Solid–Liquid Interface. *Nat. Commun.* **2014**, *5*, 4400.

(23) Cafolla, C.; Voitchovsky, K. Real-Time Tracking of Ionic Nano-Domains under Shear Flow. *Sci. Rep.* **2021**, *11* (1), 1–9.

(24) Trewby, W.; Faraudo, J.; Voïtchovsky, K. Long-Lived Ionic Nano-Domains Can Modulate the Stiffness of Soft Interfaces. *Nanoscale* **2019**, *11* (10), 4376–4384.

(25) Siretanu, I.; Ebeling, D.; Andersson, M. P.; Stipp, S. L. S.; Philipse, A.; Stuart, M. C.; van den Ende, D.; Mugele, F. Direct Observation of Ionic Structure at Solid-Liquid Interfaces: A Deep Look into the Stern Layer. *Sci. Rep.* **2015**, *4* (1), 4956–4956.

(26) Bertrand, E.; Blake, T. D.; Coninck, J. D. Influence of Solid–Liquid Interactions on Dynamic Wetting: A Molecular Dynamics Study. *J. Phys. Condens. Matter* **2009**, *21* (46), 464124.

(27) Guo, Y.; Surblys, D.; Kawagoe, Y.; Matsubara, H.; Liu, X.; Ohara, T. A Molecular Dynamics Study on the Effect of Surfactant Adsorption on Heat Transfer at a Solid-Liquid Interface. *Intl. J. Heat Mass Transf.* **2019**, *135*, 115–123.

(28) Robin, P.; Kavokine, N.; Bocquet, L. Modeling of Emergent Memory and Voltage Spiking in Ionic Transport through Angstrom-Scale Slits. *Science* **2021**, *373* (6555), 687–691.

(29) Karim, E. T.; He, M.; Salhoumi, A.; Zhigilei, L. V.; Galenko, P. K. Kinetics of Solid–Liquid Interface Motion in Molecular Dynamics and Phase-Field Models: Crystallization of Chromium and Silicon. *Philo. Trans. Royal Soc. A* **2021**, *379* (2205),

(30) Li, Z.; Ruiz, V. G.; Kanduč, M.; Dzubiella, J. Highly Heterogeneous Polarization and Solvation of Gold Nanoparticles in Aqueous Electrolytes. *ACS Nano* **2021**, *15* (8), 13155–13165.

(31) Natarajan, S. K.; Behler, J. Neural Network Molecular Dynamics Simulations of Solid–Liquid Interfaces: Water at Low-Index Copper Surfaces. *Phys. Chem. Chem. Phys.* **2016**, *18* (41), 28704–28725.

(32) Tavakol, M.; Newbold, A.; Voïtchovsky, K. Electrified Nanogaps under an AC Field: A Molecular Dynamics Study. *J. Phys. Chem. C* **2024**, *128* (49), 21050–21059.

(33) Tavakol, M.; Voïtchovsky, K. Water and Ions in Electrified Silica Nano-Pores: A Molecular Dynamics Study. *Phys. Chem. Chem. Phys.* **2024**, *26* (33), 22062–22072.

(34) Wang, J.; Li, H.; Tavakol, M.; Serva, A.; Nener, B.; Parish, G.; Salanne, M.; Warr, G. G.; Voïtchovsky, K.; Atkin, R. Ions Adsorbed at Amorphous Solid/Solution Interfaces Form Wigner Crystal-like Structures. *ACS Nano* **2024**, *18* (1), 1181–1194.

(35) Ricci, M.; Spijker, P.; Stellacci, F.; Molinari, J.-F.; Voitchovsky, K. Direct Visualization of Single Ions in the Stern Layer of Calcite. *Langmuir* **2013**, *29* (7), 2207–2216.

(36) Cafolla, C.; Bui, T.; Bao Le, T. T.; Zen, A.; Tay, W. J.; Striolo, A.; Michaelides, A.; Greenwell, H. C.; Voïtchovsky, K. Local Probing of the Nanoscale Hydration Landscape of Kaolinite Basal Facets in the Presence of Ions. *Mater. Today Phys.* **2024**, *46*, 101504.

(37) Rapaport, D. C. *The Art of Molecular Dynamics Simulation*, 2nd ed.; Cambridge University Press: Cambridge, 2004.

(38) Jorgensen, W. L. Quantum and Statistical Mechanical Studies of Liquids. 10. Transferable Intermolecular Potential Functions for Water, Alcohols, and Ethers. Application to Liquid Water. *J. Am. Chem. Soc.* **1981**, *103* (2), 335–340.

(39) Berendsen, H. J. C.; Postma, J. P. M.; van Gunsteren, W. F.; Hermans, J. Interaction Models for Water in Relation to Protein Hydration. In *Intermolecular Forces: Proc. Fourteenth Jerusalem Symp. Quantum Chem. Biochem*; Pullman, B., Ed.; Springer Netherlands: Dordrecht, 1981; pp 331–342.

(40) Jorgensen, W. L.; L, W.; Chandrasekhar, J.; Madura, J. D.; Impey, R. W.; Klein, M. L. Comparison of Simple Potential Functions for Simulating Liquid Water. *J. Chem.* **1983**, *79* (2), 926.

(41) Berendsen, H. J. C.; Grigera, J. R.; Straatsma, T. P. The Missing Term in Effective Pair Potentials. *J. Phys. Chem.* **1987**, *91* (24), 6269–6271.

(42) Bernal, J. D.; Fowler, R. H. A Theory of Water and Ionic Solution, with Particular Reference to Hydrogen and Hydroxyl Ions. *J. Chem. Phys.* **1933**, *1* (8), 515–548.

(43) Jorgensen, W. L. Revised TIPS for Simulations of Liquid Water and Aqueous Solutions. *J. Chem. Phys.* **1982**, *77* (8), 4156–4163.

(44) Horn, H. W.; Swope, W. C.; Pitera, J. W.; Madura, J. D.; Dick, T. J.; Hura, G. L.; Head-Gordon, T. Development of an Improved Four-Site Water Model for Biomolecular Simulations: TIP4P-Ew. *J. Chem. Phys.* **2004**, *120* (20), 9665–9678.

(45) Abascal, J. L. F.; Sanz, E.; García Fernández, R.; Vega, C. A Potential Model for the Study of Ices and Amorphous Water: TIP4P/Ice. *J. Chem. Phys.* **2005**, *122* (23), 234511.

(46) Abascal, J. L. F.; Vega, C. A General Purpose Model for the Condensed Phases of Water: TIP4P/2005. *J. Chem. Phys.* **2005**, *123* (23), 234505. https://doi.org/10.1063/1.2121687.

(47) Izadi, S.; Anandakrishnan, R.; Onufriev, A. V. Building Water Models: A Different Approach. *J. Phys. Chem. Lett.* **2014**, *5* (21), 3863–3871.

(48) Piana, S.; Donchev, A. G.; Robustelli, P.; Shaw, D. E. Water Dispersion Interactions Strongly Influence Simulated Structural Properties of Disordered Protein States. *J. Phys. Chem. B* **2015**, *119* (16), 5113–5123.

(49) Stillinger, F. H.; Rahman, A. Improved Simulation of Liquid Water by Molecular Dynamics. *J. Chem. Phys.* **1974**, *60* (4), 1545–1557.

(50) Mahoney, M. W.; Jorgensen, W. L. A Five-Site Model for Liquid Water and the Reproduction of the Density Anomaly by Rigid, Nonpolarizable Potential Functions. *J. Chem. Phys.* **2000**, *112* (20), 8910–8922.

(51) Rick, S. W. A Reoptimization of the Five-Site Water Potential (TIP5P) for Use with Ewald Sums. *J. Chem. Phys.* **2004**, *120* (13), 6085–6093.

(52) Nada, H.; van der Eerden, J. P. J. M. An Intermolecular Potential Model for the Simulation of Ice and Water near the Melting Point: A Six-Site Model of H2O. *J. Chem. Phys.* **2003**, *118* (16), 7401–7413.

(53) Nada, H. Anisotropy in Geometrically Rough Structure of Ice Prismatic Plane Interface during Growth: Development of a Modified Six-Site Model of H2O and a Molecular Dynamics Simulation. *J. Chem. Phys.* **2016**, *145* (24), 244706.

(54) Wu, Y.; Tepper, H. L.; Voth, G. A. Flexible Simple Point-Charge Water Model with Improved Liquid-State Properties. *J. Chem. Phys.* **2006**, *124* (2), 024503.

(55) Qiu, Y.; Nerenberg, P. S.; Head-Gordon, T.; Wang, L.-P. Systematic Optimization of Water Models Using Liquid/Vapor Surface Tension Data. *J. Phys. Chem. B* **2019**, *123* (32), 7061–7073.

(56) Fuentes-Azcatl, R.; Barbosa, M. C. Flexible Bond and Angle, FBA/$\epsilon$ Model of Water. *J. Molec. Liqu.* **2020**, *303*, 112598.

(57) Fennell, C. J.; Li, L.; Dill, K. A. Simple Liquid Models with Corrected Dielectric Constants. *J. Phys. Chem. B* **2012**, *116* (23), 6936–6944.

(58) Izadi, S.; Onufriev, A. V. Accuracy Limit of Rigid 3-Point Water Models. *J. Chem. Phys.* **2016**, *145* (7), 074501.

(59) Wang, L.-P.; Martinez, T. J.; Pande, V. S. Building Force Fields: An Automatic, Systematic, and Reproducible Approach. *J. Phys. Chem. Lett.* **2014**, *5* (11), 1885–1891.

(60) Kadaoluwa Pathirannahalage, S. P.; Meftahi, N.; Elbourne, A.; Weiss, A. C. G.; McConville, C. F.; Padua, A.; Winkler, D. A.; Costa Gomes, M.; Greaves, T. L.; Le, T. C.; Besford, Q. A.; Christofferson, A. J. Systematic Comparison of the Structural and Dynamic Properties of Commonly Used Water Models for Molecular Dynamics Simulations. *J. Chem. Inf. Model.* **2021**, *61* (9), 4521–4536.

(61) Vega, C.; Abascal, J. L. F. Simulating Water with Rigid Non-Polarizable Models: A General Perspective. *Phys. Chem. Chem. Phys.* **2011**, *13* (44), 19663–19688.

(62) Buchner, R.; Hefter, G. T.; May, P. M. Dielectric Relaxation of Aqueous NaCl Solutions. *J. Phys. Chem. A* **1999**, *103* (1), 1–9.

(63) Zasetsky, A. Yu.; Svishchev, I. M. Dielectric Response of Concentrated NaCl Aqueous Solutions: Molecular Dynamics Simulations. *J. Chem. Phys.* **2001**, *115* (3), 1448–1454.

(64) Nörtemann, K.; Hilland, J.; Kaatze, U. Dielectric Properties of Aqueous NaCl Solutions at Microwave Frequencies. *J. Phys. Chem. A* **1997**, *101* (37), 6864–6869.

(65) Chandra, A. Static Dielectric Constant of Aqueous Electrolyte Solutions: Is There Any Dynamic Contribution? *J. Chem. Phys.* **2000**, *113* (3), 903–905.

(66) Bourg, I. C.; Sposito, G. Molecular Dynamics Simulations of the Electrical Double Layer on Smectite Surfaces Contacting Concentrated Mixed Electrolyte (NaCl-CaCl2) Solutions. *J. Coll. Interf. Sci.* **2011**, *360* (2), 701–715.

(67) Welch, D. A.; Mehdi, B. L.; Hatchell, H. J.; Faller, R.; Evans, J. E.; Browning, N. D. Using Molecular Dynamics to Quantify the Electrical Double Layer and Examine the Potential for Its Direct Observation in the In-Situ TEM. *Adv. Struct. Chem. Imag.* **2015**, *1* (1), 1.

(68) Monger, H. C.; Kelly, E. F. Silica Minerals. In *Soil Mineralogy with Environmental Applications*; John Wiley & Sons, Ltd, 2002; pp 611–636.

(69) Shang, X.; Benderskii, A. V.; Eisenthal, K. B. Ultrafast Solvation Dynamics at Silica/Liquid Interfaces Probed by Time-Resolved Second Harmonic Generation. *J. Phys. Chem. B* **2001**, *105* (47), 11578–11585.

(70) Park, S.; Lee, H.; Kim, H.-E.; Jung, H.-D.; Jang, T.-S. Bifunctional Poly (l-Lactic Acid)/Hydrophobic Silica Nanocomposite Layer Coated on Magnesium Stents for Enhancing Corrosion Resistance and Endothelial Cell Responses. *Mater. Sci. Engin. C* **2021**, *127*, 112239.

(71) Rabinovich, Y. I.; Esayanur, M. S.; Johanson, K. D.; Adler, J. J.; Moudgil, B. M. Measurement of Oil-Mediated Particle Adhesion to a Silica Substrate by Atomic Force Microscopy. *J. Adhes. Sci. Technol.* **2002**, *16* (7), 887–903.

(72) Reiher, W. E. Theoretical Studies of Hydrogen Bonding. PhD, Harvard University, Cambridge, MA, 1985.

(73) Gouy, M. Sur La Constitution de La Charge Électrique à La Surface d'un Électrolyte. *J. Phys. Theor. Appl.* **1910**, *9* (1), 457–468.

(74) Chapman, D. L. A Contribution to the Theory of Electrocapillarity. *London Edinburgh Dublin Philos. Mag. J. Sci.* **1913**, *25* (148), 475–481.

(75) Emami, F. S.; Puddu, V.; Berry, R. J.; Varshney, V.; Patwardhan, S. V.; Perry, C. C.; Heinz, H. Force Field and a Surface Model Database for Silica to Simulate Interfacial Properties in Atomic Resolution. *Chem. Mater.* **2014**, *26* (8), 2647–2658.

(76) Hockney, R. W.; Eastwood, J. W. Computer Simulation Using Particles. *Computer Simulation Using Particles* **2021**.

(77) MacKerell, A. D. Jr.; Bashford, D.; Bellott, M.; Dunbrack, R. L. Jr.; Evanseck, J. D.; Field, M. J.; Fischer, S.; Gao, J.; Guo, H.; Ha, S.; Joseph-McCarthy, D.; Kuchnir, L.; Kuczera, K.; Lau, F. T. K.; Mattos, C.; Michnick, S.; Ngo, T.; Nguyen, D. T.; Prodhom, B.; Reiher, W. E.; Roux, B.; Schlenkrich, M.; Smith, J. C.; Stote, R.; Straub, J.; Watanabe, M.; Wiórkiewicz-Kuczera, J.; Yin, D.; Karplus, M. All-Atom Empirical Potential for Molecular Modeling and Dynamics Studies of Proteins. *J. Phys. Chem. B* **1998**, *102* (18), 3586–3616.

(78) Thompson, A. P.; Aktulga, H. M.; Berger, R.; Bolintineanu, D. S.; Brown, W. M.; Crozier, P. S.; in 't Veld, P. J.; Kohlmeyer, A.; Moore, S. G.; Nguyen, T. D.; Shan, R.; Stevens, M. J.; Tranchida, J.; Trott, C.; Plimpton, S. J. LAMMPS - a Flexible Simulation Tool for Particle-Based Materials Modeling at the Atomic, Meso, and Continuum Scales. *Comput. Phys. Commun.* **2022**, *271*, 108171.

(79) Stukowski, A. Visualization and Analysis of Atomistic Simulation Data with OVITO– the Open Visualization Tool. *Modelling Simul. Mater. Sci. Eng.* **2009**, *18* (1), 015012.

(80) J. D. Hunter. Matplotlib: A 2D Graphics Environment. *Comput. Sci. Engin.* **2007**, *9* (03), 90–95.

(81) Fernández, D. P.; Mulev, Y.; Goodwin, A. R. H.; Sengers, J. M. H. L. A Database for the Static Dielectric Constant of Water and Steam. *J. Phys. Chem. Ref. Data* **1995**, *24* (1), 33–70.

(82) Yagasaki, T.; Matsumoto, M.; Tanaka, H. Lennard-Jones Parameters Determined to Reproduce the Solubility of NaCl and KCl in SPC/E, TIP3P, and TIP4P/2005 Water. *J. Chem. Theory Comput.* **2020**, *16* (4), 2460–2473.

(83) Huang, J.; Rauscher, S.; Nawrocki, G.; Ran, T.; Feig, M.; de Groot, B. L.; Grubmüller, H.; MacKerell, A. D. CHARMM36m: An Improved Force Field for Folded and Intrinsically Disordered Proteins. *Nat. Methods* **2017**, *14* (1), 71–73.

(84) Beglov, D.; Roux, B. Finite Representation of an Infinite Bulk System: Solvent Boundary Potential for Computer Simulations. *J. Chem. Phys.* **1994**, *100* (12), 9050–9063.

(85) Lewis-Atwell, T.; Townsend, P. A.; Grayson, M. N. Comparisons of Different Force Fields in Conformational Analysis and Searching of Organic Molecules: A Review. *Tetrahedron* **2021**, *79*, 131865.

(86) Tavakol, M.; Vaughan, T. J. Energy Dissipation of Osteopontin at a HAp Mineral Interface: Implications for Bone Biomechanics. *Biophys. J.* **2022**, *121* (2), 228–236.

# Graphical abstract

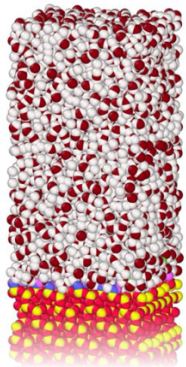 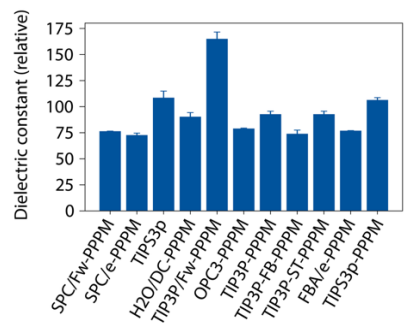

# Supplementary Information

# Comparative molecular dynamics simulations of charged solid-liquid interfaces with different water models

Mahdi Tavakol[1,*] and Kislon Voïtchovsky[1,*]

1. Physics Department, Durham University, Durham DH1 3LE, UK

*corresponding authors: mahditavakol90@gmail.com, kislon.voitchovsky@durham.ac.uk

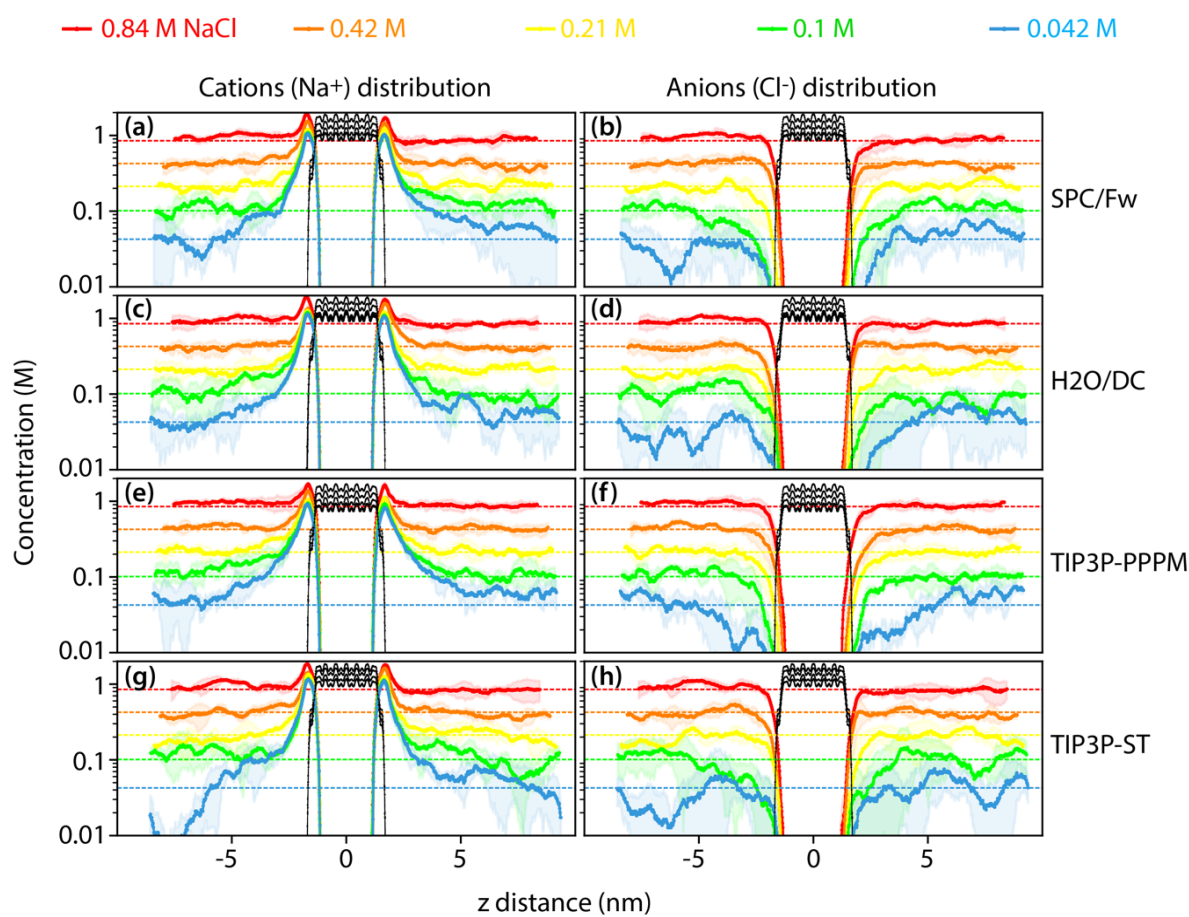

*Figure S1*: Effect of the salt concentration on the ion distribution near solid-liquid interface for the selected water models. The cation and anion distribution near a solvated silica surface is shown for SPC/Fw (a) and (b), respectively, $H_2O$/DC (c) and (d), TIPS3p-PPPM (e) and (f), and TIP3P-(g) and (h). The different colours represent the different bulk ionic concentrations (horizontal dashed line) with the silica density appearing in black.

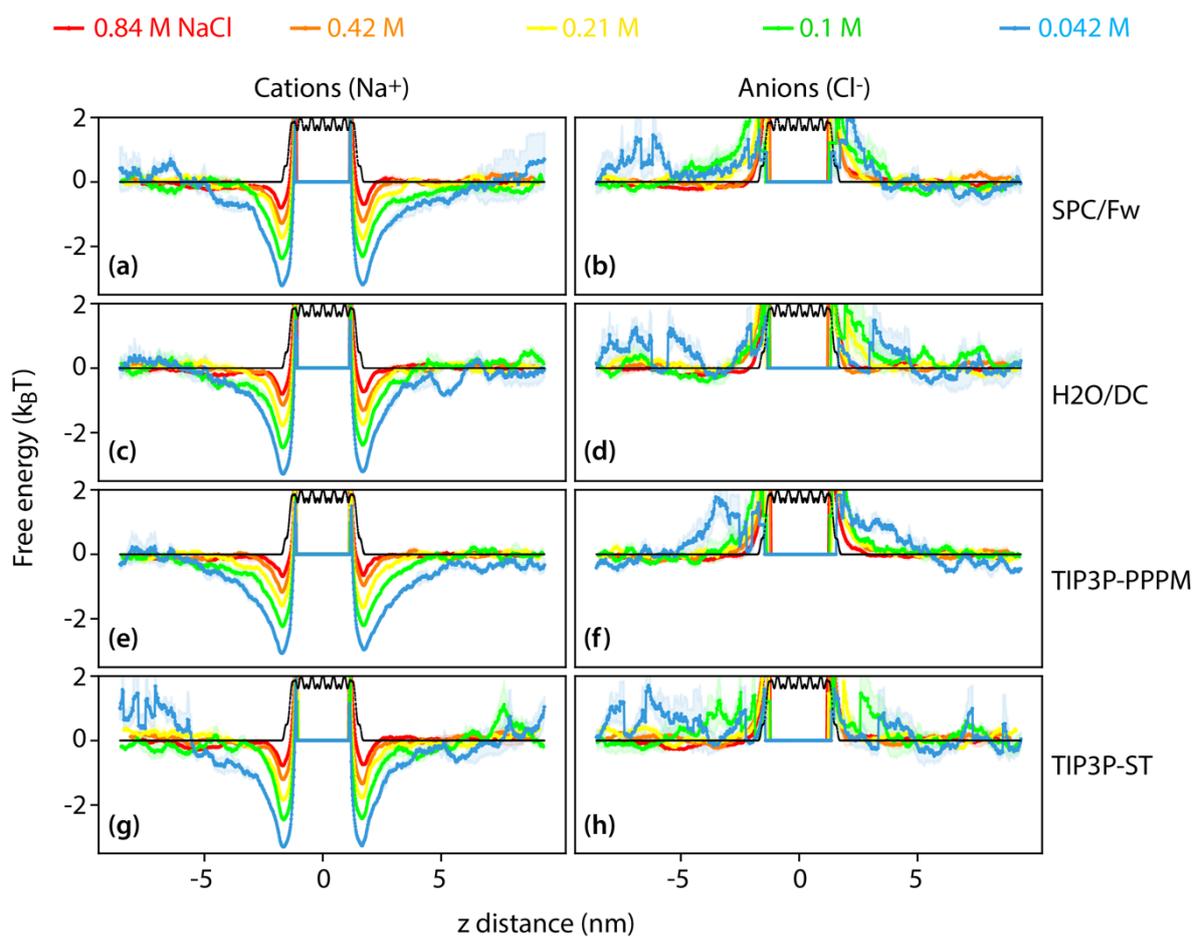

*Figure S2*: Effect of the salt concentration on the free energy of the ions near solid-liquid interface for the selected water models. The cation and anion free energy distribution near a solvated silica surface is shown for SPC/Fw (a) and (b), respectively, H$_2$O/DC (c) and (d), TIPS3p-PPPM (e) and (f), and TIP3P-(g) and (h). The different colours represent the different bulk ionic concentrations with the silica density appearing in black.

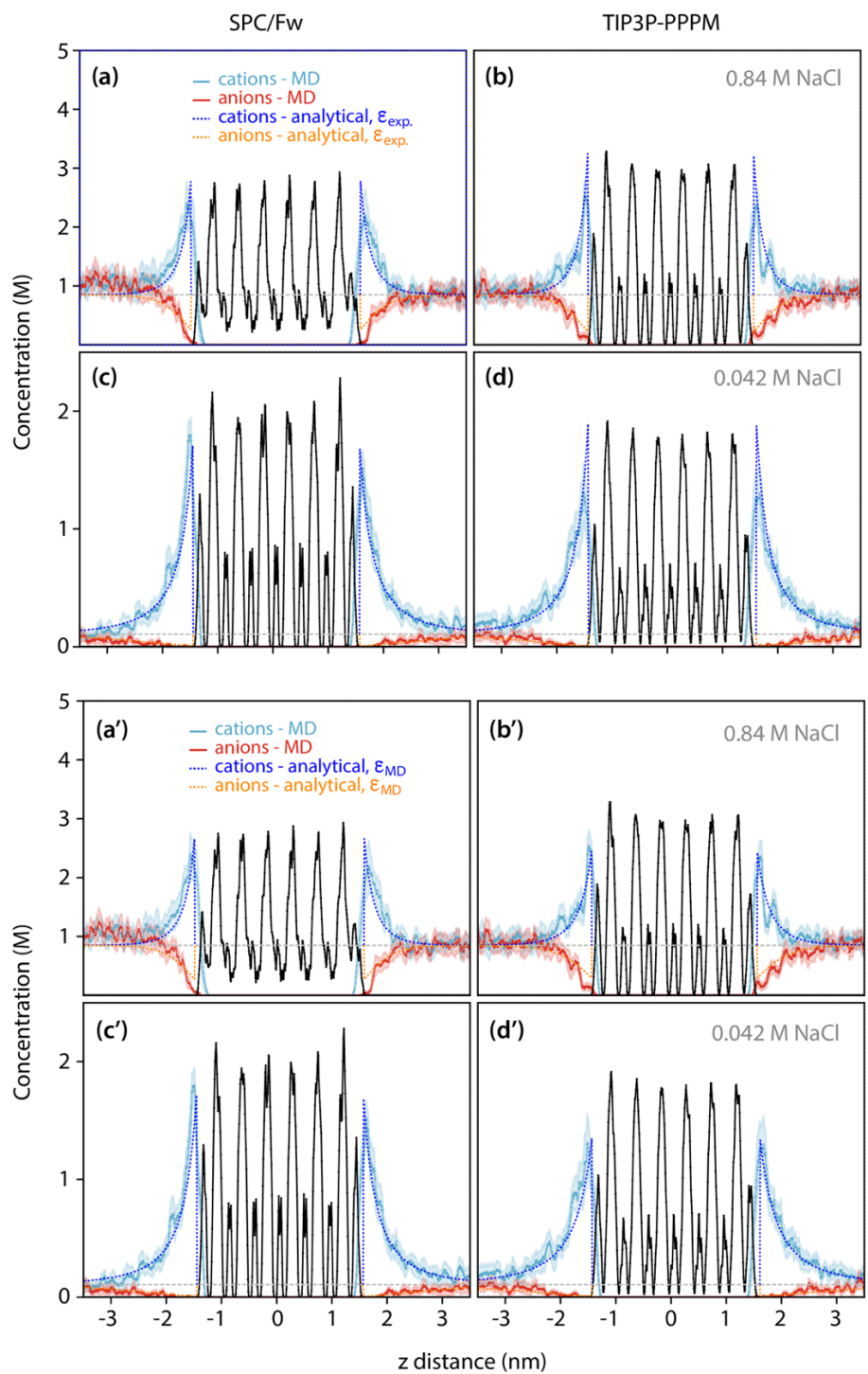

*Figure S3*: Ion density profiles obtained with SPC/Fw and TIPS3p-PPPM at two salt concentrations. For comparison, analytical predictions from Gouy-Chapman model are given using the with experimentally measured dielectric constant for water $\varepsilon_{exp.}$ (a-d) and with the constant $\varepsilon_{MD}$ derived from MD simulations (a'-d'). The SPC/Fw water model tends to underestimates the counter-ion density near the silica surface at 0.84 M, and effect more marked when the MD calculated dielectric constant is used. However, the predictions made by TIPS3p-PPPM water are in a good agreement with the analytical solution despite its poor ability to predict the dielectric constant accurately. The counter-ion and co-ion concentration profiles near the charged silica surface are given at 0.84 mM NaCl and 0.042 M NaCl for SPC/Fw with analytical predictions based on $\varepsilon_{exp.}$ (a, c) and $\varepsilon_{MD}$ (a', c') respectively. The counter-ion and co-ion concentration profiles at 0.84 mM NaCl and 0.042 M NaCl for TIPS3p-PPPM with analytical predictions based on $\varepsilon_{exp.}$ (b, d) and $\varepsilon_{MD}$ (b', d') respectively.

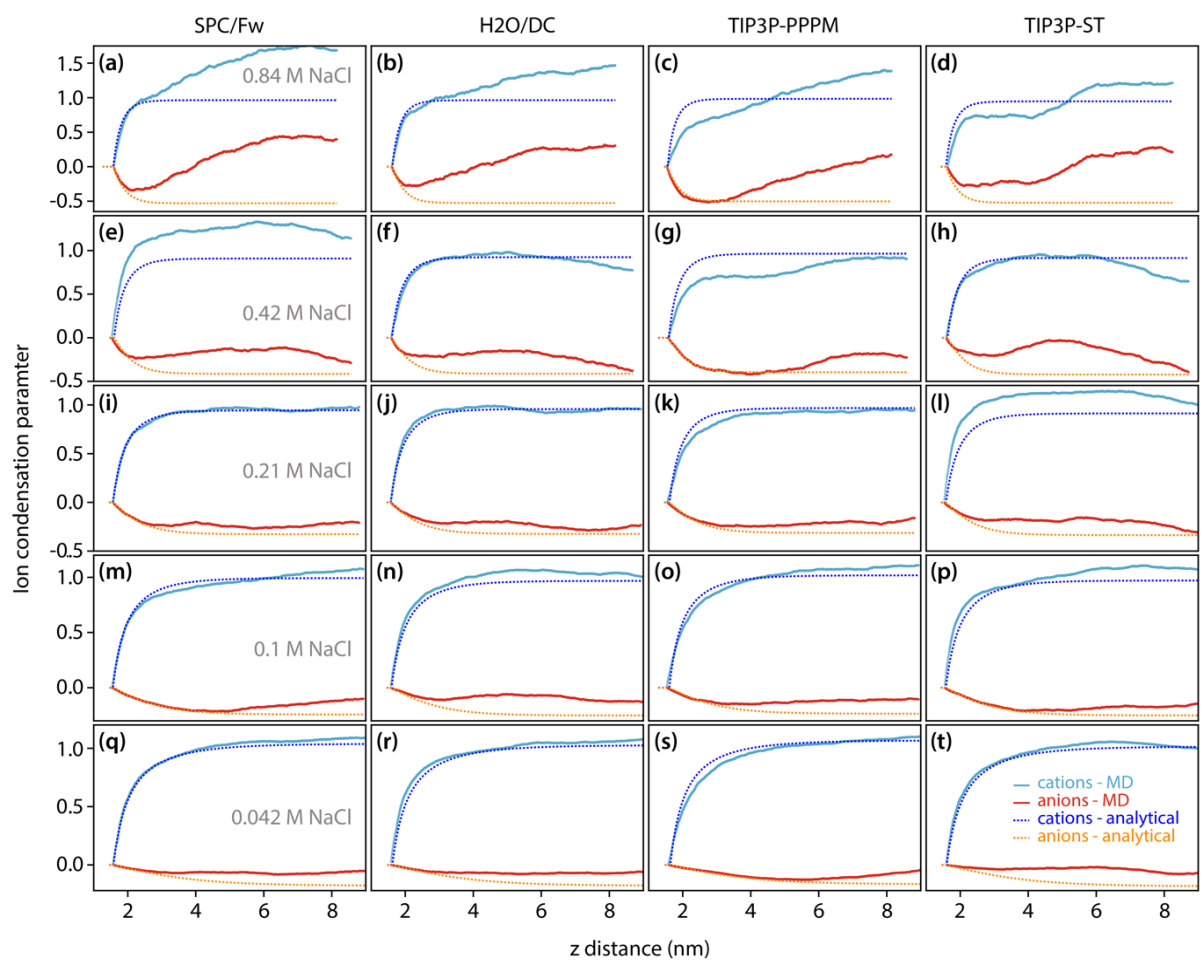

*Figure S4*: Comparison of the condensation parameter obtained at various salt concentrations for the selected water models. The results are compared with analytical prediction from the Gouy-Chapman model. Counter-ion (blue) and co-ion (red) condensation profiles are shown with a solid line for the MD simulations and a dashed line for the analytical predictions. The SPC/Fw -derived condensation parameter is given for 0.84 M NaCl (a), 0.42 M NaCl (e), 0.21 M NaCl (i), 0.1 M NaCl (m) and 0.042 M NaCl (q) respectively. The H2O/DC -derived condensation parameter is given for 0.84 M (b), 0.42 M (f), 0.21 M NaCl (j), 0.1 M (n) and 0.042 M (r) respectively. The TIPS3p-PPPM -derived condensation parameter is given for 0.84 M (c), 0.42 M (g), 0.21 M NaCl (k), 0.1 M (o) and 0.042 M (s) respectively. The TIPS3P-ST -derived condensation parameter is given for 0.84 M (d), 0.42 M (h), 0.21 M NaCl (l), 0.1 M (p) and 0.042 M (t) respectively.